\documentclass{aa}
\usepackage{graphicx}
\begin{document}

\title{The nature of the absorbing torus in compact Radio Galaxies
}
\author{G. Risaliti\inst{1,2}, L. Woltjer\inst{1,3}, M. Salvati\inst{1}
}

\institute{
INAF - Osservatorio Astrofisico di Arcetri, Largo E. Fermi 5, I-50125 Firenze, Italy
\and
Harvard-Smithsonian Center for Astrophysics, 60 Garden Street,
Cambridge, MA, 02138, USA
\and
Observatoire de Haute-Provence, CNRS, 04870 Saint-Michel l'Observatoire, France
}
\authorrunning{G. Risaliti et al.}
\titlerunning{Absorber in Radio Galaxies}
\offprints{}

\date{Received / Accepted}

\abstract{ We present BeppoSAX observations of the two radio galaxies PKS 1934-63 and S5 1946+708.
Strong Iron K$\alpha$ lines are detected in both objects indicating that the
two sources are absorbed by column densities higher than  
10$^{24}$ cm$^{-2}$. 

Combining radio continuum, HI absorption and X-ray data we can constrain the physical state and the
dimensions of the absorber. We find that the putative obscuring torus is molecular and located
at a distance higher than 20 pc from the center in S5 1946+70. In PKS 1934-63
no radio nucleus has been observed. If this is due to free-free absorption a
radius of only a few pc is inferred. Since the K$\alpha$ lines have been
detected only at 90\% confidence, we also discuss the implications if they
would be not as strong as found from our data. From our results it appears
that the combination of hard X-ray data and spatially resolved HI absorption
measurements provides a powerful diagnostic for conditions in the absorbing
``torus''.
}
\maketitle

%In one case, NGC 6300, the line profile is clearly asymmetric to the red and broad, and is best fitted
%with a relativistic profile.


\section{Introduction}                                                                           

In unified models of Seyfert galaxies Sy1/Sy2 and of quasars/radio galaxies Q/RG the Broad Line
Region BLR and the optical nucleus are hidden at certain orientations by an ``absorbing torus''.
Direct evidence for such a torus comes from absorption at the lower energies X-rays from
the nucleus.  While the "torus" needs not have a truly toroidal shape, it has to be geometrically thick.  For
the Seyferts the observed ratio of Sy1/Sy2 (1/3-1/6) suggests that the torus as seen from the
 nucleus subtends an angle of order  $\pi/2$, and corresponding estimates have been made for RG.

Polarized optical broad emission lines in some Sy2 and RG have been ascribed to material above
the torus, which scatters nuclear light towards us.  Also X-rays may be scattered by relatively
cool material with the iron K$\alpha$ line at 6.4 keV appearing in emission.  When column densities in
the torus approach $N_{\rm H} = 10^{24}$ cm$^{-2}$, the direct X-rays are much attenuated and the equivalent
width (EW) of the iron line may become 1 keV or more. If the scattering material is warm
iron lines at 6.7-6.9 keV may appear.

While the X-ray data yield values or limits for N$_{\rm H}$, they do not contain information on the
physical and kinematical conditions in the torus.  Absorption at 21 cm by atomic hydrogen
(HI) may be informative in this respect.  HI absorption has been detected mainly in powerful
compact radio galaxies.  Line widths of up to nearly 1000 km s$^{-1}$ have been observed,
indicating that the HI is located rather close to the nucleus, presumably in the torus.
In other cases the velocities are smaller and the location of the HI more uncertain.

The absorbing matter generally covers only a small part of the radio source.
It is therefore essential to have high resolution, preferably VLBI, data to determine the optical
depth in HI in front of the nucleus.  Conway \& Blanco (1995) obtained VLA data for
CygA and found that the X-ray and HI data may be quantitatively fitted to simple models
of the torus.  They also concluded that the radius of the torus should exceed 15 pc if
strong free-free absorption is to be avoided, which would make the radio nucleus unobservable
at 1.4 GHz.  This radius is larger than inferred for typical Seyferts.

To obtain further information on the tori in RG we have observed the X-ray spectra of two
RG with Beppo-SAX. PKS 1934-63 is a powerful compact double with a separation of 158 pc.
VLBI observation failed to locate a nucleus between these components even at 8 GHz
(Tzioumis et al. 1999).  It has been suggested that very strong free-free absorption might
be responsible (Woltjer 2000).  HI measurements (V\'eron-Cetty et al. 2000) also failed to find
HI absorption in front of the radio source, except for a narrow weak feature, presumably due
to an isolated cloud.  The second source observed with BeppoSAX is S5~1946+708, a compact symmetric
object (CSO\footnote{Compact symmetric objects are defined as lobe-dominated 
sources smaller than 1 kpc in overall size (Conway 2002)} 
with a nucleus and two jets and an overall dimension of 100 pc.
The HI absorption profile in front of the nucleus has a total width of 600-800 km s$^{-1}$ and
the absorption in front of the jets shows that the projected thickness of the ``torus'' is about
30 pc with some narrower absorption further out (Peck et al. 1999).  Further data
on the two sources are given in Table 1. Everywhere we have taken H$_0$= 67
km s$^{-1}$ Mpc$^{-1}$\\

\section{Observations and analysis of the BeppoSAX data}

PKS 1934-63 was observed on Nov 11/12, 2000 and again on May 03-04 2001, each time for about
50 ksec.  While in May the K$\alpha$ line appeared to have a lower EW than before, the low S/N ratio
was insufficient to conclusively demonstrate variability and so we combined the two data sets.  The total
exposure times were 38 ksec for the LECS instrument (0.1-10 keV), 95 ksec for
the MECS (1.65 - 10 keV) and 48 ksec for the PDS (20-200 keV).
The source was not detected with the PDS.
The LECS and MECS images show the presence of an unknown serendipitous source 5 arcmin NW
 of the target, with comparable flux.
To avoid contamination, the spectrum of PKS 1934-63 was extracted in a circular region with
radius 2 arcmin using the XSELECT reduction package. This small radius is the one for which 
we obtain the highest S/N. The spectral analysis has been performed
with the XSPEC code.
The LECS and MECS spectra have been convolved with the response matrices provided by the
BeppoSAX Science Data Center and background subtracted on the basis of a long exposure of a
blank field.  To check for systematic errors in this procedure we also extracted the background
spectrum from a source free region in our field. After rescaling for the vignetting factor the
 two spectra agreed very well.

\begin{table}
\centerline{\begin{tabular}{lcc}
\hline
\hline
&PKS 1934-63&S5 1946+708 \\
\hline
z&0.182&0.101\\
S$_{1.4 GHz}$(Jy) & 15 & 1.0 \\
$\log P_{1.4 GHz}$(W Hz$^{-1}$) & 27.1 & 25.3 \\
A$_V$(mag) & 1.5 & 2.2 \\
$\log L_{\rm [OIII]}$ (erg s$^{-1}$) & 43.2 & 42.0 \\
\hline
\end{tabular}}
\caption{\footnotesize{Characteristics of the two radio galaxies. Subsequent
lines give the redshift, the 1.4 GHz flux density from the nucleus, the radio power, the
visual absorption for the Narrow Line Region and the absorption corrected
[OIII]$\lambda$5007 luminosity (H$_0=67$~km s$^{-1}$ Mpc$^{-1}$).}}
\end{table}

Given the low S/N we can only fit simple models.  A power law fit to the continuum yields a photon
index $\Gamma$=1.9, but leaves a clear excess at 5.3 keV corresponding to 6.3 keV in the rest frame.
This is interpreted as a Fe K$\alpha$ line with very large EW=2.0 keV. 
If we interpret this spectrum as reflection-dominated, we can assume a
reflection efficiency, R, to estimate the intrinsic continuum and a lower limit
for the absorbing column density, N$_{\rm H}$. Assuming R=5\% (the maximum allowed
according to reflection models as in Ghisellini et al. 1994, and therefore
the one giving the lowest intrinsic continuum) we find N$_{\rm H}>2.5\times 10^{24}$
cm$^{-2}$. 
The results 
are given in Table 2.
This interpretation is not unique, due to the poor signal-to-noise.

An alternative possibility is suggested by the separate analysis of the two observations.
The 2-10 keV continuum is higher in the second observation, while the line flux remains
constant. We fitted the two spectra with the photon indexes and the line flux fixed at the
values obtained in the combined fit, leaving the power law normalization free.
We found that the continuum variation is significant at a confidence level of 90\%.
In the high-continuum observation the line is 
detected only at a 1.2$\sigma$ level.

Therefore, we cannot exclude that the observed spectrum is the intrinsic emission from
the active nucleus, but that a variation by a factor of $\sim$2 in the continuum (not followed
by a line variation, indicating that the line is produced by 
material far from the center) gives rise to a measured equivalent width in the combined
spectrum higher than the usual values for type 1 AGNs (100-300 eV).

The spectra from the two observations are plotted in Fig. 1. The results of the
separate fits are listed in Table 2.

\begin{figure}
\centerline{\resizebox{\hsize}{!}
{\includegraphics{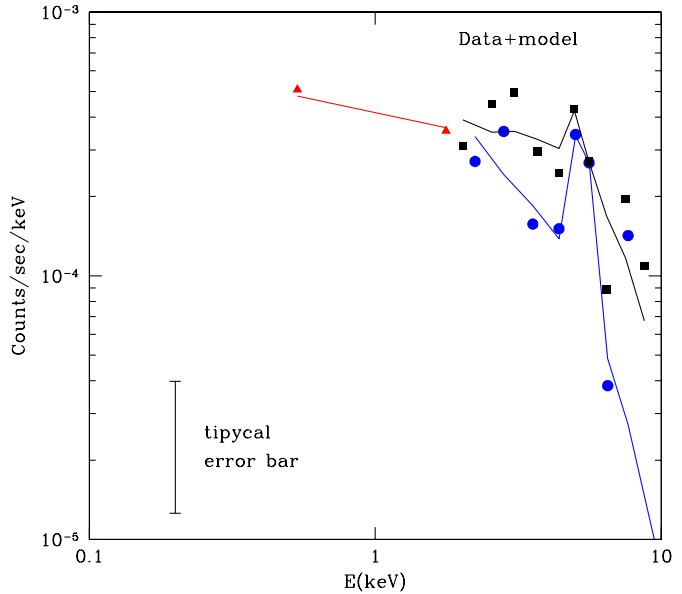}}}
\caption{\label{fig:s1}
\footnotesize{Spectrum and best fit model for the BeppoSAX
observation of PKS 1934-63. Triangles: LECS data of the first observation; circles:
MECS data of the first observation; squares: MECS data of the second observation. }}
\end{figure}

\begin{table*}
\centerline{\begin{tabular}{lccc}
\multicolumn{4}{c}{PKS 1934-63} \\
\hline
\hline
Parameter & Merged Obs. & First obs. & Second obs. \\
\hline
Power law: &&&\\
Photon index & $1.9^{+0.5}_{-0.6}$ & 2.2$_{-0.6}^{+0.5}$ & 1.2$_{-0.7}^{+0.7}$ \\
Normalization (10$^{-5}$ cts s$^{-1}$ keV$^{-1}$) & 3.8$^{+3.2}_{-2.0}$& 4.4$^{+4.3}_{-2.4}$& 
2.2$_{-1.5}^{+3.0}$ \\
\hline
Iron line: &&& \\
Peak energy (keV, rest frame) & 6.26$^{+0.34}_{-0.38}$ & 6.31$_{-0.21}^{+0.29}$ &
6.4$^a$\\
Equivalent width (keV) & 2$^{+1.2}_{-1.4}$ & 3.2$_{-2.6}^{+3.1}$ & 0.6$_{-0.6}^{+0.9}$ \\
\hline
Flux 2-10 keV (10$^{-13}$ erg cm$^{-2}$ s$^{-1}$)$^b$& 1.5$\pm 0.2$ & 1.2$\pm 0.2$ &
2.3$\pm 0.3$\\
$\chi^2$/d.o.f. & 14.2/17& 4.3/7 & 4.4/7\\
\hline
\end{tabular}}
\caption{\footnotesize{Best fit models for the merged observations of PKS 1934-63 (first column)
and for the two observations taken separately (second and third column). From the non-detection
in the PDS we can estimate N$_{\rm H}>2.5\times10^{24}$ cm$^{-2}$, assuming a reflection
efficiency of 5\%. Notes: $^a$: fixed parameter, $^b$: uncorrected for
absorption.}}
\end{table*}
 
S5 1946+708 was observed by BeppoSAX for 40 ksec.  Again there is another source in the field
 at 10 arcmin from our target. 
%It has been identified as a cataclismic
%variable (Wei
%et al. 1999). 
The PDS flux detected in the field could come from either source (see the
discussion in the Appendix)
and so we have no conclusive evidence for hard radiation from S5 1946+708.
Fitting the spectrum as before we obtain from the LECS and MECS data: $\Gamma$=2.6 and a line at 6.9
keV in the rest frame with EW = 1.8 keV  (Table 3 and Fig. 2). 
With a 90\%-confidence lower limit at 6.6 keV this would suggest reflection by warm matter.
However, if we were to fix the energy of the line at 6.4 keV, its statistical significance would
not change much. We can also obtain a lower
limit for the absorbing column density, N$_{\rm H}$, assuming that the PDS emission
is entirely due our target, as we did for the best fit shown in Table 2. 
However, in the
opposite hypothesis (i.e. PDS emission entirely due to the serendipitous
source) the lower limit on N$_{\rm H}$ increases, but the other parameters do
not significantly change.
Finally, a low-energy cut-off is detected,
indicating a second absorber along the line of sight of the reflected
component. 

\begin{figure}
\centerline{\resizebox{\hsize}{!}
{\includegraphics{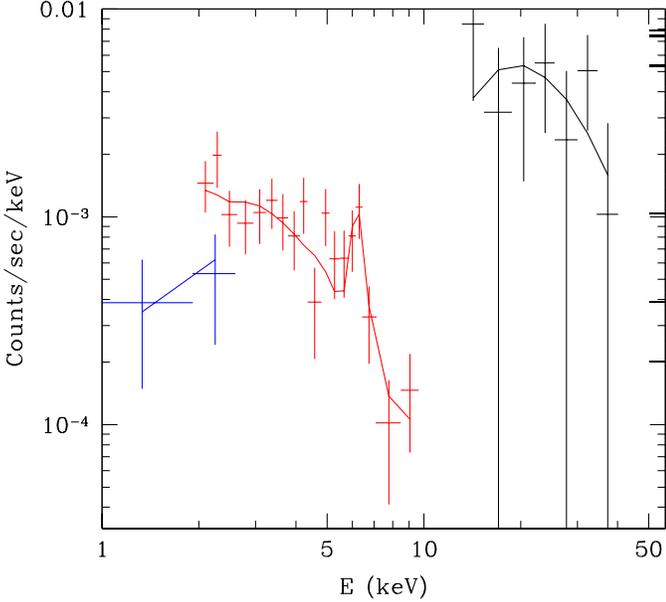}}}
\caption{\label{fig:s2}
\footnotesize{Spectrum and best fit model for the BeppoSAX
observation of S5 1946+708.}}
\end{figure}

\begin{table}
\centerline{\begin{tabular}{lc}
\multicolumn{2}{c}{S5 1946+708} \\
\hline
\hline
Parameter & Best fit value \\
\hline
Power law: &\\
Photon index & $\Gamma=2.6^{+0.6}_{-0.7}$ \\
Normalization & 5.4$^{+10}_{-3.7} 10^{-4}$ cts s$^{-1}$ keV$^{-1}$ \\
high N$_{\rm H}$ absorber & N$_{\rm H1}>2.8\times 10^{24}$ cm$^{-2}$ \\
low N$_{\rm H}$ absorber & N$_{\rm H2}=2.6_{-1.9}^{+2.5}\times 10^{22}$ cm$^{-2}$ \\
\hline
Iron line: & \\
Peak energy (rest frame) & E=6.87$^{+0.19}_{-0.24}$ keV \\
Equivalent width & EW=1.8$^{+1.2}_{-1.1}$ keV \\
\hline
Flux 2-10 keV (10$^{-13}$ erg cm$^{-2}$ s$^{-1})^b$& 5.3$\pm 0.6$ \\
$\chi^2$/d.o.f. & 16.5/21\\
\hline
\end{tabular}}
\caption{\footnotesize{Best fit model for S5 1946+708. The high column density
absorber, N$_{\rm H1}$, has been estimated assuming that all the
PDS emission is due to this source. $^b$: uncorrected for absorption.}}
\end{table}

In conclusion both objects have rather steep spectra.
Combined with the high EW of the iron lines this suggests that the sources are Compton thick
(log N$_{\rm H} > 24$), with the observed X-rays resulting from scattering.
Observations with higher S/N ratio are needed to confirm these results.\\

\section{Discussion} 

We adopt a simple model in which a torus with radius r$_{\rm pc}$ parsec is illuminated by a point source
nucleus emitting X-rays with an energy spectral index $\alpha$ = 0.7 between 1 and 100 keV with a
luminosity of 10$^{44}$ L$_{44}$ erg s$^{-1}$.  The torus is taken to be isobaric with 
temperature 10$^3$ T$_3$ K and
particle density 10$^6$ n$_6$ cm$^{-3}$.  We neglect the curvature of the torus. 
Our aim is to use radio and X-ray data to constrain the density and size of
the circumnuclear absorber.

We shall make use of the
calculations of Maloney (1996) and Neufeld et al. (1994) for single phase isobaric models at 
high pressure ($p/k\sim~10^{11}$ K~cm$^{-3}$), and of Maloney et al. (1996) at low pressure
($p/k\sim~10^{9}$ K~cm$^{-3}$).
Conditions in the torus depend on the effective ionization parameter  $\xi_{\rm eff}$ 
which from Maloney et al.
we write, for  $\alpha$ =-0.7, as 
\begin{equation}
\xi_{\rm eff}=0.17 L_{44} N_{24}^{-0.9} n_6^{-1} r_{\rm pc}^{-2}
\end{equation}

with N$_{24}$ the column
density towards the nucleus in units of 10$^{24}$ cm$^{-2}$ and $r_{\rm pc}$ the
distance to the center in pc. Different values of $\alpha$ should not change
the results too much.

All tori have an atomic zone towards the nucleus, but if the total column density through the
torus \~{N}$_{24}$ is large enough this is followed by a mainly molecular zone.  From the calculations of
Neufeld et al. and of Maloney it is seen that the transition from atomic to molecular occurs in a
very  narrow region from log$\xi_{\rm eff}\sim -2.0$ when the gas is fully atomic to $\log \xi_{\rm eff}=-3.4$
when it is fully molecular, for pressure \~{p} = p/k in the range
10$^{10} - 10^{12}$ K cm$^{-3}$. In the high pressure range the temperature averages about 7000 K 
in the atomic zone and some 600 K in the molecular zone. 
From Maloney et al. (1996) we see that at lower pressures \~{p} in the range
 10$^7$ - 10$^9$ K cm$^{-3}$ the transition is more gradual and temperatures as low as 100 K may be reached 
in the atomic zone. 
In a typical atomic zone the average
ionized fraction is around log(n$_e$/n)$\sim$-1.8 and may be lower at pressures near 10$^7$ K cm$^{-3}$.
In the molecular zone n$_e$ is negligible.

The condition for a fully atomic torus, log$\xi_{\rm eff} > -2.0$, can be rewritten as
a condition in the torus radius versus density plane:
\begin{equation}
\log n_6+2 \log r_{\rm pc} < log L_{44}-0.9 \log \tilde{N}_{24}+1.3
\end{equation}

where \~N is the total column density to the outer edge of the torus.

The ratio n$_{\rm HI}$/n in the molecular zone depends on the destruction of H$_2$ molecules by X-rays and
on their formation on the surface of grains.  Maloney et al. obtain an expression that in the
molecular zone may be written as  
\begin{equation}
\log(n_{\rm HI}/n)=3.08+\log\xi_{\rm eff}
\end{equation}
where the constant involves
uncertain parameters relating to the grains.  It should be noted that all these calculations depend
on still somewhat uncertain chemistry at low temperatures, on the degree of trapping of IR lines
and on the nature of the grains.  Maiolino et al. (2000) have suggested that grains in the torus
 may be different from those in the local interstellar medium or even perhaps largely absent.

An additional condition may be that in a steady state the pressure in the torus should be higher than
or equal to the radiation pressure (Maloney 1996). 
This can be written as follows:
 
\begin{equation}
\tilde{p}_{11}>2 f_{\rm abs} L_{44} r_{\rm pc}^{-2}
\end{equation}
with \~{p}=10$^{11}$ \~{p}$_{11}$
and $f_{\rm abs}$ the fraction of the X-rays absorbed in the torus.
In case only gas pressure is relevant we can obtain a condition in the
radius-density plane:

\begin{equation}
\log n_6+2 \log r_{\rm pc} > log L_{44}+\log f_{\rm abs} - log T_3 + 2.3 + w
\end{equation}

where $w=0$ if the gas is atomic and $w=0.3$ if molecular.
We are now in a position to use our observations to test the hypothesis that the absorbing torus is 
atomic.
\subsection{Compton-thick tori}
The X-ray spectrum yields the total column density \~{N}$_{\rm H}$. When strong relatively narrow iron
lines with EW around 1 keV or more are seen, the torus is likely to be Compton thick with \~{N}$_{24} > 1$
(Maiolino et al. 1998).  In this case the observed X-rays have been reflected or scattered and the intrinsic L$_{\rm X}$
should exceed the observed value in the 2-10 keV range by a substantial factor $\zeta$. 
We conclude that both of our sources
have \~{N}$_{24}>1$ and we shall conservatively adopt $\zeta$=50. With this assumption, we
have L$_{44}$=5 for both sources. Taking T$_3$=7 in an atomic torus, 
\~N$_{24}$=1 and for the Compton thick case $f_{\rm abs}=1$,
we see that Eqs. (2) and (5) are incompatible.
We note that this result is independent of L$_{\rm X}$. Thus a high pressure Compton thick torus
cannot be atomic if Eq. (4) holds.
\~{N}$_{24}=1$ is a lower limit to the actual value of
N$_{\rm H}$. However, for higher values the discrepancy
between Eqs. (2) and (3) increases further.

For a molecular torus $\log \xi_{\rm eff} < -3.4$ and Eq. (5) is trivially
satisfied.

%The pressure condition for a molecular gas is obtained from Eq. (1),
%assuming T$_3$=0.6\footnote{for the rest of the discussion we assume
%$f_{\rm abs}=1$}:
%\begin{equation}
%\log n_6+2 \log r_{\rm pc} > \log L_{44} - \log T_3 + 3.08
%\end{equation}
%
%This equation is represented by the solid steep line in Fig. 3, where we
%assumed L$_{44}$=5 (corresponding to $\xi$=10).

We can now complement the information on the total column density coming from
the X-ray data with the radio data, which provide constraints on the 
free-free absorption, $\tau_{\rm ff}$, and the column density of atomic hydrogen
\~N$_{\rm HI}$.
  
%At 1.4 GHz we have from V\'eron Cetty et al. (2000) for uniform atomic torus
%\begin{equation}
%\tau_{\rm ff} = 1.7 \times 10^6  q^2 T_3^{-3/2} n_6 \tilde{N}_{24}
%\end{equation}
%with q the ionized fraction n$_e$/n.
  
If the torus is mainly molecular
there is still an atomic zone on the inside, which is responsible for the free-free opacity.
Neufeld et al. (1994) give an expression for $\tau_{\rm ff}$
which at 1.4 GHz corresponds to  
\begin{equation}
\tau_{\rm ff} \sim  270 L^{1.1}_{44} \tilde{p}_{11}^{-0.1} r_{\rm pc}^{-2}
\end{equation}

The contribution of the molecular zone to $\tau_{\rm ff}$ is negligible.

%In the following we compare the radio and X-ray data for our two sources with
%the cases for molecular or atomic tori, in the Compton-thin and Compton-thick
%scenarios.
 
Since the nucleus of S5 1946+708 has been observed at 1.4 GHz, the free-free
opacity must satisfy $\tau_{\rm ff} \leq 3$. 
From Eq. (6) we then obtain:
\begin{equation}
0.1 \log n_6+2 \log r_{\rm pc} > 2.13+1.1 \log L_{44}
\end{equation}

This inequality is plotted in Fig. 3.

Peck et al. (1999) have measured the HI absorption in front of the
nucleus of the source, obtaining for the HI column density log \~{N}$_{\rm HI}$ = 22.5 + log T$_{\rm sp,3}$,
with T$_{\rm sp}$ the spin temperature.

At high densities radiative
excitation is negligible and in the molecular zone we should have
T$_{\rm sp}$=T$\sim$600 K, or log\~{N}$_{\rm HI}$=22.3 which for \~{N}$_{24}=1$ 
requires log(n$_{\rm HI}$/n)=-1.7, or from Eq. (3) log$\xi_{\rm eff}$=- 4.8. 

Treating the torus as a uniform layer we should evaluate 
$\xi_{\rm eff}$ 
at the mid point N$_{24}$=0.5.
From Eq. (1), we then obtain:

\begin{equation}
\log n_6+2\log r_{\rm pc}=5.0
\end{equation}

This corresponds to a line in the $n_6$-$r_{\rm pc}$ plot. Taking into account also
the condition on the free-free opacity, only the part of the line with $r_{\rm pc}$
larger than given by Eq. (7), corresponds to allowable configurations. So
the minimum radius of the torus is about 22 pc. Since Peck et al. found
evidence for free-free absorption we should expect the actual value to be not
very far from the minimum, corresponding to $n_6$ around 200. This is also
suggested by the half thickness of the HI absorption of only 15 pc.

%With L$_{44}=\xi(L_{44})_{\rm obs~2-10 keV}=39\times0.13=5.1$ 
%we then have $\xi_{\rm eff}=1.62 n_6^{-1} r_{\rm pc}^{-2}$ and
%therefore $n_6 r^2_{\rm pc}=0.98\times 10^5$.  Eq. (3) yields $n_6 r^2_{\rm pc}>0.17\times 10^{5}$
%and therefore the pressure
%condition is satisfied.  The condition $\tau_{\rm ff} < 3$ (Eq. (5)) now becomes r $> 23 P_{11}^{-0.05}$ pc 
%and therefore
%n$_6 < 180$ upon neglecting the P$_{11}^{0.05}$ factor.  
%In fact with n$_6$=180 P$_{11}=1.1$.  From the fact that
%Peck et al. (1999) found evidence for free-free absorption P and n$_6$ are likely to be close to these
%limits.  It is easily verified that in this case the atomic zone ends (at log$\xi_{\rm eff}=-2$) at
%N$_{\rm HI,24}=0.0054$ and so contributes only negligibly (2\%) to the HI optical depth.

%Computing $\tau_{\rm ff}$ explicitly from Eq. (4) yields $\tau_{\rm ff}=1.7$ in semiquantitative agreement with the
%result from Eq. (5).  Evidently all our results are quantitatively still somewhat uncertain because
%of the use of a two zone model, rather than a full integration through the torus.  However the
%uncertainties in the chemistry in the molecular zone do not justify a more accurate treatment.
%Nevertheless it appears that a high pressure (p$_{II}\sim1$) molecular torus with r$>$23 pc 
%is compatible
%with the  data on S5 1946+70, while a low pressure atomic torus presents serious difficulties.

With $n_6=200$ and \~{N}$_{24}=$1 the path length through the dense gas   
is only 0.002 pc.
It is therefore likely that such a torus would be composed of small high density clouds or filaments.
As stated above, the models being used here are single phase at any given
radius. If the absorber is made by dense clouds, there is probably a confining
medium which could be relevant in the absorption processes we are studying.
We note however that a medium with temperature T$> 10^8$~K and
density $n_6\sim10^{-3}$ would have the right pressure to
confine our clouds, without significantly contributing either to the radio
or to the X-ray absorption. A denser, colder (T$< 10^7$~K, $n_6\sim0.1)$
confining gas could instead be relevant. 
However, this possibility can be ruled out, for it would imply a
very high thermal emission in the soft X-rays, which is not observed.

{All the above discussion is focused to the high pressure case. It is worth 
considering the possibility of a low pressure, low temperature atomic torus.
Assuming \~p$\sim 10^9$~K~cm$^{3}$ and $T_3\sim 0.1$ we still have a
relatively high density. This implies that the spin temperature is 
$\sim$ equal to the thermal temperature. Therefore, the relation of Peck. et
al. 1999 rules out the possibility of a Compton thick torus. We will further discuss
this scenario in the case of a Compton-thin torus (Sect. 3.2).}

%The alternative interpretation of the spectrum is that the source is Compton
%thin and that the K$\alpha$ equivalent width has been overestimated or has
%been exagerated by variability. In this case the column density $\log
%N_H=22.5$. 
%In this case the position of the source in the diagnostic diagram in Fig. 3
%becomes about the standard for unobscured Seyfert galaxies.
%
%The HI absorption (Peck et al. 1989) indicates a column density in
%HI equal to $\log N_{HI}=19.5+\log T_{\rm sp}$. Since in the atomic torus T$\sim$
%8000 K, we would have $\log N_{HI}=23.4$, higher than the total $N_H$. This
%rules out the case for an atomic, Compton thin torus.
%
%On the other hand, a molecular torus could be envisaged but the required
%pressure is rather high. From Eq. (7) with $\tau_{\rm ff}<3$ we have
%approximately r$>7$ pc. The torus can only be molecular on its outer edge when
%$\xi_{\rm eff}<10^{-3}$ or $n_6 r_{\rm pc}^2>2\times 10^3$. For r=7 pc we have n$_6>40$.
%With $T_3=0.6$ in the molecular zone we would have p$_{11}>0.24$ which is to
%be compared with $p_{11}=4\times 10^{-4}$ in
%an atomic torus with the same radius.

{\bf PKS 1934-63:} Less information is available for PKS 1934-63.
%From the large EW of the Fe K$\alpha$ line we again
%infer the source to be Compton thick and therefore N$_{24}$ at least equal to unity. The same assumptions
%as before give $\xi$=39 and L$_{44}$=4.7. 
Since no nucleus has been observed at 8 GHz we infer 
$\tau_{\rm ff}>3$
at that frequency, corresponding to $\tau_{\rm ff}>98$ at 1.4 GHz. However since
the dynamic range in the 8 GHz observations is not very high this condition
may be a bit too strict.
From Eq. (7) we then obtain
\begin{equation}
0.1 \log n_6+2 \log r_{\rm pc} < 0.66+1.1\log L_{44}
\end{equation}

%For an atomic torus with N$_{24}$=1 Eq. 4
%then yields n$_6>4.3$. From Eq. 1 we have at N$_{24}$=1,  log$\xi_{\rm eff}<-0.1\sim$ log(n$_6 r_{\rm pc}^2$).
%To have an atomic torus log$\xi_{\rm eff}>-2$ and hence log(n$_6 r^2_{\rm pc})<1.9$. But from the pressure
%condition (Eq. 3) we have log (n$_6 r^2_{\rm pc})>3.3$ and so we again run into trouble. At the adopted
%X-ray luminosity, the atomic torus has inadequate pressure to balance the radiation pressure.

The permitted parameter space for a molecular torus in PKS 1934-63 is therefore the shaded region
at the bottom-right of Fig. 3b. The distance of the absorber
must be lower than $\sim$ 4~pc.

The case for an atomic torus in this source can be easily ruled out, using the
same equations as above to show that the condition $\log \xi_{\rm eff}>-2$ and Eq.
(8) are not compatible.

Some comments are in order. The adopted models are all based on the
assumption that the tori in the two sources are Compton thick. We have taken
\~N$_{\rm H}$=1, though the rather uncertain PDS results suggest perhaps even higher
values. However as far as we are aware no detailed calculations have been made
for tori with $N_{24}>1$. While a thicker torus would not change $\tau_{\rm ff}$,
the inferred N$_{\rm HI}$ might be affected. A second point is that the values of
L$_{44}$ are very uncertain if the sources are indeed Compton thick and might
well be larger. And thirdly the basic assumption has been made that the same
medium is responsible for the X-ray, free-free and HI absorption.
\subsection{Compton thin tori}
From our rather limited X-ray data we have concluded that the EW of K$\alpha$
is large in both sources. However in several cases (e.g. 3C 120, Zdziarski \& Grandi 2001) the
K$\alpha$ line EW has much diminished as better observations and more
sophisticated modelling of the continuum were implemented. While our present
data are insufficient for the construction of more detailed models, the fact that the
strong K$\alpha$ is at less than 90\% confidence, indicates that it might be
worthwhile to investigate what would happen if the K$\alpha$ EW would
come down to values more appropriate for Compton thin sources, 
this the more so since also variability may have marked effects on the
EW (see e. g. 3C 382, Grandi et al. 2001). In the case of S5 1946+708 we then
would conclude from Table 3 that, in fact, $\log \tilde{N}_{\rm H}=22.5$. From the
data of Peck et al. we have $\log \tilde{N}_{\rm HI}=22.5+\log T_{\rm sp,3}$ and thus
for a high pressure atomic torus $\log N_{\rm HI}=23.4$ which is evidently impossible. So the
torus should be molecular. With $T_{\rm sp,3}=0.6$ we have $\log \tilde{N}_{\rm HI}=22.3$ and
$\log n_{\rm HI}/n =-0.2$, corresponding from Eq. (3) to $\log \xi_{\rm eff}=-3.3$
at the mid point of the torus where $\log N_{\rm H}=-1.8$. We have $\zeta=4$ since
the absorption is small and $\log L_{44}$=-0.2. From Eq. (1) we then find
$\log(n_6 r_{\rm pc}^2)=3.95$ (line labelled ``N$_{24}=0.03$'' in Fig. 3a.

At lower pressures a cool atomic torus becomes a possibility. Since from the
data of Peck et al. $\log \tilde{N}_{\rm HI}=22.5+\log T_{\rm sp,3}$ and since now
$\log \tilde{N}_{\rm H}=22.5$ we should have $T_{\rm sp,3}=1$. According to the results
of Maloney et al. (1996) for $n_6=0.1$ this requires $\log \xi_{\rm eff}=-2.0$ and
for $n_6=0.001$, $\log \xi_{\rm eff}=-1.25$. Again taking a one zone model where
these values refer to the midpoint $\log N_{\rm H}=22.2$ we obtain from Eq. (1) that
for $n_6=0.1$ $r=67$ pc and for $n_6=0.001$, $r=282$ pc. From the $\log n -
\log r$ plot we see that connecting the two points an extrapolation to higher
densities should fit to the line for the molecular torus, which, in fact, is
only marginally molecular. The pressures now are very modest: for $n_6=0.1$ we
have $p_{11}=0.001$, a value more representative for the NLR. It may also be
verified that $\tau_{\rm ff}$ is of the order of a few times 0.1. 

%For a molecular torus on the other hand we find from Eq. (5) r$<3.9 P_{11}^{-0.05}$ pc. From Eq. 3
%we have n$_6 r^2_{\rm pc}>15700$ and hence from Eq. 1 at N$_{24}$=0.5  that log$\xi_{\rm eff}<4.02$ 
%amply meeting
%the condition for a molecular torus.

%From r$<$ 3.9 pc we then have n$_6>1030$ and hence P$_{11}>6.2$.  So the inferred strong free-free
%absorption in PKS 1934-63 is the consequence of the small radius of the torus and its high pressure.
%With in the NLR pressures of P$_{11}=10^{-3}-10^{-2}$ and perhaps up to 1 in extreme cases and in 
%the BLR with \~p$_{II}\sim$10-1000 or even more the pressures found in
%the torus in between the two do not seem unreasonable.

%We illustrate the constraints inferred in this discussion in Fig. 4.
\begin{figure}
\centerline{\resizebox{\hsize}{!}
{\includegraphics{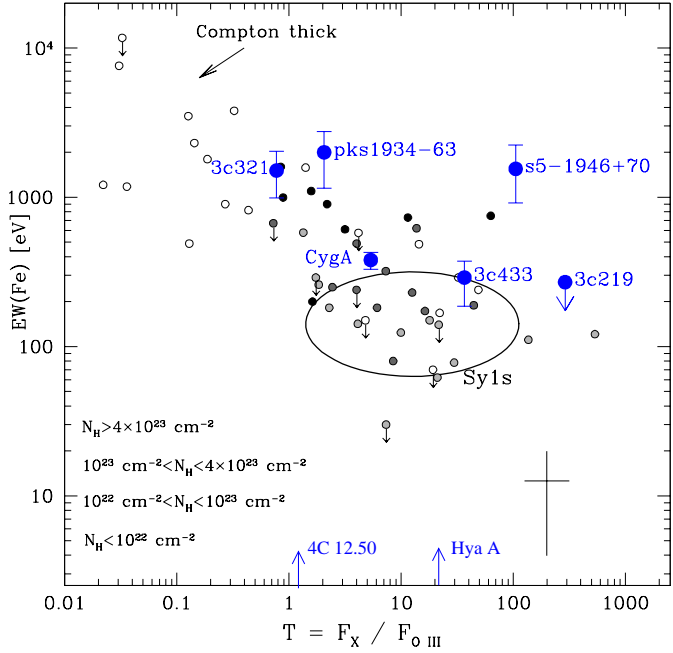}}}
\caption{\label{fig:s3}
\footnotesize{Diagnostic diagram for Seyfert 2s of Bassani et al.
(1999). We have superimposed the data for a sample of powerful radio galaxies for which
a hard X-ray observation and [OIII] and Balmer lines measurements are
available. Our two sources lie significantly above the region occupied by
both Seyfert 2s and Compton-thin radio loud objects. }}
\end{figure}

We now analyze the case for Compton-thin absorption in PKS 1934-63. A low energy cut off is
not required by our BeppoSAX observations. From our data we estimate N$_{\rm H}<2\times10^{22}$ cm$^{-2}$.
Assuming N$_{\rm H}=2\times10^{22}$ cm$^{-2}$ we have log L$_{44}$=-0.09 and $f_{\rm abs}=0.1$.

With a temperature T=7000 K, the pressure condition for an atomic torus gives:

\begin{equation}
\log n_6 +2\log r_{\rm pc} > 1.36
\end{equation}

while the condition $\xi_{\rm eff}>-2$ gives:

\begin{equation}
\log n_6 +2\log r_{\rm pc} < 2.66
\end{equation}

and with the condition $\tau_{\rm ff}>98$ and Eq. (6) we have $\log n_6 >2.3$ and
$r<1.5$~pc.

Therefore in this case an atomic torus is acceptable.

Note however that the pressure condition (Eq. (5)) is not satisfied. Therefore,
this scenario is possible only if the absorbing gas is not in a steady state,
or other mechanisms contribute to the internal pressure, in addition to the
thermal component (for example, magnetic fields).

%We now can complement this result with the constraint on $\tau_{\rm ff}$.
%
%At 1.4 GHz we have from V\'eron Cetty et al. (2000) for a uniform atomic torus
%\begin{equation}
%\tau_{\rm ff} = 1.7 \times 10^6  q^2 T_3^{-3/2} n_6 \tilde{N}_{24}
%\end{equation}
%with q the ionized fraction n$_e$/n.
\begin{figure}
\centerline{\resizebox{\hsize}{!}
{\includegraphics{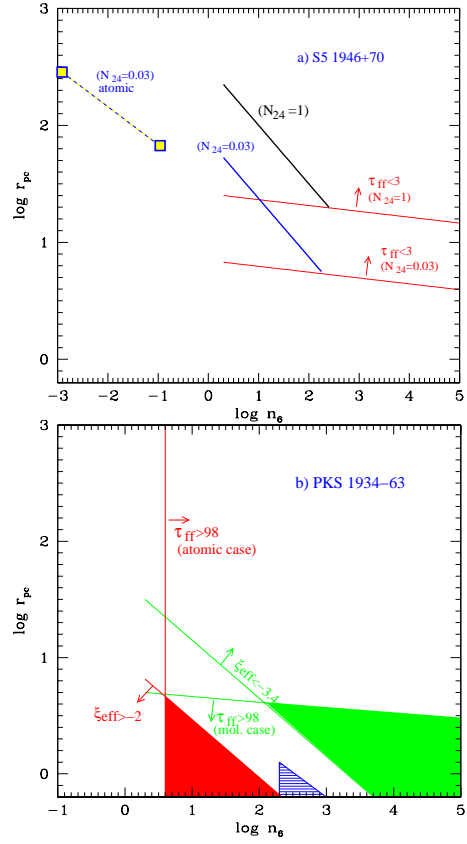}}}
\caption{\label{fig:s4}
\footnotesize{Radius-density diagram for the torus around our radio
galaxies. {\bf a)} S5 1946+708: the permitted parameters are given by the relation between
total absorbing column density, HI column density and spin temperature 
(Peck et al. 1999). The two almost horizontal lines represent the constraints
on the free-free opacity. The two points on the left represent a solution with
an atomic, Compton thin torus, in the two cases of $n_6=0.1$ and $n_6=0.001$.
The dotted lines connecting these
points represent a possible solution in the low-density, low pressure regime.
{\bf b)} PKS 1934-63: In the molecular torus case, the allowed region is at
the bottom-right of the diagram, and is delimited by the constraints on the
ionization parameter $\xi_{\rm eff}<-3.4$, and on the free-free opacity
$\tau_{\rm ff}>98$. In the atomic torus case, the condition on the ionization
parameter becomes $\xi_{\rm eff}>-2$, while the opacity conditions gives only a
constraint on the minimum density. The permitted region (the filled triangle) 
is however by a large factor 
incompatible with the requirement for the gas pressure to be equal to or
greater than the radiation pressure. The small hatched triangle is obtained in
the same way, but assuming a column density $N_{\rm H}=3\times10^{22}$ cm$^{-2}$.
}}
\end{figure}

%In a typical atomic zone the average
%ionized fraction is around log(n$_e$/n)$\sim$-1.8.

%From $\tau_{\rm ff}>98$ we then obtain $\log n_6 > 2.32$ and from Eq. 11
%$r<1.5$~pc.
The permitted region in the radius-density plane is indicated in Fig. 3b. The
pressure is high, $p_{11}\sim 1$, but should be possible at such a small
radius.

Some other powerful radio galaxies have data in the literature. Cyg A was
discussed by Conway \& Blanco (1995) who fit their HI data with an atomic
torus with $n_6<0.2$ and $r_{\rm pc}>15$. For a molecular torus Maloney (1996)
finds $r_{\rm pc}>50$.

For Hya A Taylor (1996) found HI absorption corresponding to $\log
N_{\rm HI}=23.1+\log T_{\rm sp,3}$, while Sambruna et al. (2000) detected an absorbed
X-ray nucleus with $\log N_{\rm H}=22.5$ and $\log L(2-10 {\rm keV})=42.2$ (for
$H_0=67$ km s$^{-1}$ Mpc$^{-1}$). To reconcile the values of $N_{\rm H}$ and
$N_{\rm HI}$ we have to take $T_{\rm sp}$=250 K which excludes a high pressure atomic
torus with T=8000 K. However from the results of Maloney et al. (1996) at
lower pressures we find a fit for $n_6$=0.1 and T=250 K with $\log
\xi_{\rm eff}=-2.5$ at the mid point and $r_{\rm pc}$=20. It should be noted, however,
that iron K$\alpha$ line has not been measured in this source.  

%Perhaps of particular
%interest in the conext of our sources is the radio galaxy 3C 321 which has a
%$K\alpha$ line in its X-ray spectrum with an EW of 1.5 keV according to
%Sambruna et al.

Bassani et al. (1999) have presented an interesting diagram for Seyfert galaxies with as coordinates
the equivalent width of the iron lines near 6.5 keV and the ratio of the observed 2-10 keV X-ray
flux to the absorption corrected [0III] $\lambda 5007$ flux.

In this diagram (Fig. 4) the Sy2 follow a broad sequence from high $F_{\rm X}/F_{\rm [OIII]}$ and low iron EW
to low $F_{\rm X}/F_{\rm [OIII]}$ and high iron EW. The interpretation of this diagram is that as the X-ray
absorption increases, the much weaker scattered component with its fluorescent iron lines becomes
more conspicuous.  Plotting in this diagram the powerful narrow line radio galaxies with adequate data there
is perhaps a tendency for the average to be displaced towards higher $F_{\rm X}/F_{[0III]}$ values,
with S5 1946+708 the most extreme case. An interesting possibility is that, since our results
point to a far absorber, the narrow line clouds emitting the [OIII] line are also
partially covered by the X-ray absorber. In this case, using the true [OIII] flux would shift
the source to the left in the plot, in the standard region for Compton-thick AGNs.
If on the other hand the K$\alpha$ line is not as strong as we found, then S5
1946+708 would come down vertically and fit in with other low absorption
objects.

Since in typical Seyferts the ionizing spectrum appears to have a strong ``blue bump'' which
is probably absent in Liners (like S5 1946+708), thereby causing an unusually low ionization,
it may well be that its absence also causes $F_{\rm X}/F_{[0III]}$ to be unusually high.
Alternatively since these very compact radio galaxies have been shown in some cases to be very
young (thousands of years, Conway 2002) it may be that the NLR has not yet had the time to fully develop.
More high quality X-ray spectra are needed to further investigate these possibilities.

\section{Concluding remarks}

Observations with BeppoSAX seem to show that both PKS 1934-63 and S5 1946+708
are Compton thick sources. This is suggested by the high equivalent width of
the iron K$\alpha$ lines (EW$>1$~keV) and, in the case of   PKS 1934-63, by
the low X-ray to [OIII] flux ratio. 
 
Powerful radio galaxies with X-ray absorbing column densities higher than 10$^{24}$
cm$^{-2}$ are expected on the basis of unified models, 
but are still rather elusive: to our knowledge, 
only one such object, 3C 321 (Sambruna et al. 1999)
has been observed prior to the present work.

We have used a simple modelization of the absorbing torus (under the
hypothesis that the same medium is responsible for both the radio and X-ray
absorption) to estimate the density and distance of the nuclear absorber.
We find that in PKS 1964-93 the absorber is compact (R$< 4$~pc) while in S5
1946+708 the distance from the center is higher than $\sim 20$~pc.

There are however some problems with the interpretation of the X-ray data as
due to Compton-thick sources.

For PKS 1934-63 this does not pose particular
problems and, in fact, could serve as an explanation of the absence of a radio
core in this compact double. However there remains the difficulty that at 90\%
confidence the source is variable, which seems in contradiction to it being
Compton thick on the basis of the high Fe K$\alpha$ line equivalent width,
also with 90\% confidence.

For S5 1946+708 there are two problems: the inferred pressure at several tens
of pc from the nucleus seems to be anomalously high, while its position in the
Bassani et al. diagram suggests a rather low X-ray absorption. Both problems
disappear when the source is not Compton thick.

As a consequence of these ambiguities, in our Discussion we have also estimated the
physical parameters of the absorbers in the scenario of Compton-thin X-ray
absorption. 
%Most narrow line radio galaxies are not Compton thick (e.g. Sambruna et al.
%1999). However these authors find 3C 321 to have an iron line with EW=1.5 keV.
%In this case strong absorption is also suggested by its position in the
%Bassani et al. diagram. Moreover it is a strong IR source most likely due to
%the reradiation in the IR of the absorbed X-rays.

For most of the radio galaxies the torus appears to have a radius of several
tens of pc. Another case where this was suggested on the basis of the radio
data alone is PKS 1353-341 (V\'eron-Cetty et al. 2000). Smaller tori as in the
Seyfert galaxies tend to be optically thick in free-free absorption (O'Dea et
al. 2000), while those in radio galaxies must in most cases be optically thin
since the radio nucleus is observable.

To make further progress higher quality X-ray spectra are needed, while also
further VLBI data on HI absorption are required.

\acknowledgements
This work was partially supported by the Italian Ministry for University and
Research (MURST) under grant Cofin-00-02-36.
\appendix
\section{Serendipitous sources}
In the field of S5 1946+708 there is another source at 10' distance identified
as the ROSAT source 1RXS J194639.7+704552. Wei et al. (1999) have identified
this source with a star (B=16.9, R=15.2) at 7'' distance just within the 8''
ROSAT error circle. It has an optical spectrum with some absorption features
and H$\alpha$ and H$\beta$ in emission. They note another star at 9' (B=15.0,
R=13.4) which seems to us an equally valid candidate, and state that it is a
cataclysmic variable. We have been unable to locate this variable in the
literature. The X-ray spectrum of this source is extremely hard
($\Gamma=1.1_{-0.2}^{+0.1}$, with F(2-10 keV)=2.6$\times 10^{-12}$ erg
cm$^{-2}$ s$^{-1}$). The low energy flux is in agreement with the ROSAT 
measurement. This source may very well account for the flux measured with the
PDS which has a field of view of 50 arcmin (Fig. 5). If so it might further increase
the likelihood that S5 1946+708 is not Compton thick. It would be useful to
have an optical spectrum of the ROSAT source. 
\begin{figure}
\centerline{\resizebox{\hsize}{!}
{\includegraphics{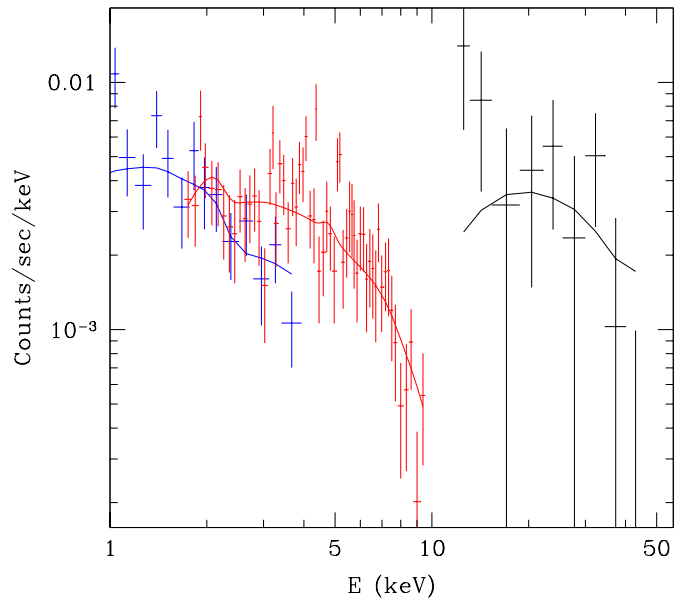}}}
\caption{\label{fig:s5}
\footnotesize{Spectrum of the serendipitous source in the field of S5 1946+708}}
\end{figure}

In the field of PKS 1934-63 there is another source at 5' distance at
$\alpha(2000)=$19 38 56.7 and $\delta(2000)=$ -63 38  2.7. This source is not identified in the
literature. With a power law fit we obtain $\Gamma=1.7_{-0.8}^{+0.7}$ 
and F(2-10 keV)=1.5$\times 10^{-12}$ erg
cm$^{-2}$ s$^{-1}$. 

\clearpage

\end{document}